# Standing wave vs Green's function approach to the Casimir force problem

Frédéric Schuller[1], Renaud Savalle[1] and Michael Neumann-Spallart[2]

[1] Observatoire de Paris, 5 Pl Jules Janssen, 92195 Meudon, France. Tel : +33 (0)1 45 07 75 86. Fax : +33 (0)9 58 41 31 41. E-mail : renaud.savalle@obspm.fr

[2] Groupe d'Etude de la Matière Condensée, CNRS/Université de Versailles, 45, av. des Etats-Unis, 78035 Versailles CEDEX, France. E-mail: mns@cnrs-bellevue.fr

*Abstract:* After a short recall of our previous standing wave approach to the Casimir force problem, we consider Lifshitz's temperature Green's function method and its virtues from a physical point of view. Using his formula, specialized for perfectly reflecting mirrors, we present a quantitative discussion of the temperature effect on the attractive force.

*Introduction*

The effect of the quantum nature of the electro-magnetic field manifests itself if retarded potentials between atoms and molecules are considered, as shown by Casimir and Polder [1]. An even more striking quantum effect with no classical analogue is the attractive force between perfectly reflecting parallel plates at zero temperature. For this effect Casimir derived in 1948 the expression [2]

(1) $\quad F_z = \dfrac{\pi^2}{240} \dfrac{\hbar c}{a^4}$

with the plates located at z = 0 and z = a.

In his derivation Casimir considers the zero point energy inside the cavity, and he obtains the force as represented by its derivative with respect to the distance a.

Here a difficulty appears given the fact that if the zero-point energy density

(2) $\quad \varepsilon = \dfrac{1}{2}\hbar\omega_k$

is summed over all possible modes **k**, a meaningless infinite result is obtained. Casimir has solved the problem by demonstrating how unphysical infinities disappear by introducing compensations from the surroundings outside the cavity. However, this delicate problem has given rise to numerous subsequent studies for which we refer to K.A. Milton's book [3]. Note that some of them are based on intricate mathematical extrapolation methods.

Another way for computing the Casimir force consists in writing down an expression derived directly from Maxwell's stress tensor instead of differentiating the energy relation. Unfortunately this method does not remove the divergences brought about by summation over modes so that here again the compensation scheme remains an essential element of the theory. This approach has been initiated by Lifshitz [4] for the more general case of dielectric plates, which then reduces to that of perfect mirrors if the dielectric constants of the plates are taken to infinity. Divergence is removed by subtracting from the electro-magnetic state inside the cavity the one that would exist inside the cavity volume if the boundaries were absent.

In a recent paper [5] we have introduced Casimir's original zero point energy situation into a stress tensor formalism. Here we shall give a short outline of that method followed by a recall of Lifshitz's results. In the latter case the temperature effect on the force has been evaluated. Note that throughout this article we assume parallel perfectly reflecting plates located at $z = 0$ and $z = a$.

*The standing wave calculation*

In [5] we consider the following expression for the field inside the cavity:.

$$(3) \quad \begin{aligned} E_x &= A_x \cos(k_x x) \sin(k_y y) \sin(k_z z) \\ E_y &= A_y \sin(k_x x) \cos(k_y y) \sin(k_z z) \\ E_z &= A_z \sin(k_x x) \sin(k_y y) \cos(k_z z) \end{aligned} \qquad \mathbf{k} = \begin{pmatrix} k_x \\ k_y \\ k_z \end{pmatrix} = \begin{pmatrix} \pi n_x / L \\ \pi n_y / L \\ \pi n_z / a \end{pmatrix}$$

with a the distance between the square plates, L their extension in the x,y directions and $n_x, n_y, n_z$ integers taking all positive values. In this way one ensures that the field obeys the boundary conditions $E_{||} = 0$ at the boundaries. Moreover eq.'s (3) satisfy the zero charge condition $\nabla \cdot \mathbf{E} = 0$.

Note that using these modes amounts to assuming that, if squared, they replace the expectation value $\langle \hat{\mathbf{E}}^2 \rangle$ of the square of the corresponding field operator $\hat{\mathbf{E}}$.

For the force exerted on the plate we introduce Maxwell's stress tensor involving both the **E** and the **B** field, with for the latter the expression

$$(4) \quad \mathbf{B} = -\frac{1}{i\omega} \nabla \times \mathbf{E}$$

We then obtain after some straightforward calculations the result

$$(5) \quad \sigma_{zz} = \varepsilon_0 E_z^2 - \frac{1}{2}\left(\varepsilon_0 \mathbf{E}^2 + \frac{1}{\mu_0}\mathbf{B}^2\right) = -\frac{1}{8}\varepsilon_0 A^2 \frac{k_z^2}{k^2}$$

Here $\sigma_{zz}$ represents the relevant tensor element which, summed over modes, yields the force per unit area on the plates.

Adjusting now the average over the square of these fields, according to the corresponding energy relation, to the zero point energy density, we arrive at the expression

$$(6) \quad A^2 = A_x^2 + A_y^2 + A_z^2 = \frac{2}{\varepsilon_0}\frac{\hbar\omega_k}{L^2 a}$$

Note that this equation has been obtained by replacing in the averages over the field squares the quantities $\sin^2$, $\cos^2$ by 1/2.

In order to perform a summation over modes we now need a frequency cut-off. Following Fierz [6] we introduce for this purpose a convergence factor $e^{-\lambda k}$. Furthermore, letting L go to infinity, we make the replacements

$$(7) \quad \sum_{n_x n_y} \to \frac{L^2}{\pi^2}\int dk_x dk_y \quad .$$

Here we indicate only the result of this straightforward calculation. Integrating and summing over all positive integers $n_z$ yields for the attractive force the expression

$$(8) \quad F_z = -\frac{\hbar c}{\pi^2}\lambda^{-4} - \frac{\hbar c}{a^4}\pi^2 \frac{6}{2\times 4!}B_4 \quad \text{with } B_4 = -1/30 \text{ a Bernoulli number.}$$

Naturally the result depends on the cut-off parameter $\lambda$ and for $\lambda \to 0$ it becomes infinite as it should, whereas the second term yields Casimir's result as given by eq. (1). We emphasize the fact that the divergent term does not depend on the distance a between the plates so that one may safely assume that it is compensated, as mentioned in the introduction. However, the way in which this occurs in detail is irrelevant in the context of our present derivation.

*The Lifshitz calculation*

The weak point of any approach that starts from the zero point energy, in the spirit of Casimir's initial work, is that it considers an ideal situation never encountered in nature. On the opposite, Lifshitz allows for finite temperatures and recovers Casimir's result as a purely mathematical limit for $T \to 0$. Note however that it is remarkable that both approaches, which at first sight might seem different from a physical point of view, yield exactly the same result. Moreover, Lifshitz's theory offers as a byproduct a discussion of the temperature effect on the force.

The Lifshitz theory is summarized in the appendix for the special case of interest here, i.e. perfectly reflecting plates. It uses temperature Green's functions which differ from the familiar Green's functions of the electro-magnetic field by the fact that they involve pseudo frequencies $\varsigma_s$ defined by the relation

(9) $\varsigma_s = \dfrac{2\pi k_B s T}{\hbar}$

with $k_B$ the Boltzmann constant and s a positive integer. Given the fact that the Casimir force is very weak in any practical case, we consider here the ratio R between the finite temperature result and the $T \to 0$ Casimir limit of the force.

Using the formulae given in the appendix with $T \to k_B T$, and setting

(10) $\dfrac{k_B}{\hbar c} = \kappa = 4.36 \times 10^2 \, SI$

we obtain for this ratio the expression

(11) $R = \dfrac{480}{\pi^3} \kappa a T \sum_{s\,1}^{\infty} \int (2\pi s \kappa a T p)^3 \dfrac{dp}{p\left(e^{4\pi s \kappa a T p} - 1\right)}$

which reduces to unity in the limit $T \to 0$ as shown in the appendix.

By means of this expression the quantity R has been evaluated numerically in order to determine the effect of temperature on the force. The figure below shows results for different values of the plate distance a. As one can see the force decreases with increasing temperature. This is understandable given the fact that the force involves the difference between the electro- magnetic field inside the cavity and the free field and that this difference should be less the higher the temperature.

Note that according to eq. (11) the quantity R depends on the product aT , meaning that the curves in the figure are related to each other by a horizontal scaling factor.

In any case, applying the Lifshitz formula, no value of R superior to unity ever occurs.

*Discussion*

Lifshitz's calculation has been criticized by Hargreaves [7] who stated « that it yet be desirable that the general theory be reexamined and maybe set up anew ». This is because Lifshitz derived an approximate expression for the temperature effect, which in our notation takes the form $R = 1 - \dfrac{16}{3}(\kappa a T)^4$. Considering the case $a = 5 \times 10^{-6}$ m he concludes that this result is not valid for this distance. That is true, since applying the formula one obtains $R = 1.2 \times 10^{-2}$ for T = 100 K which is in total disagreement with our numerical calculations. Moreover, the formula exhibits a zero initial slope not appearing on our graphs. However, this does not mean that Lifshitz's theory « is in error ».

We are not commenting here on theories which allow values of R much larger than unity [8,9]. In contrast to Lifshitz's calculations these theories rely on extreme conditions. Their physical reality demands a detailed discussion which lies beyond the scope of this article.

*Conclusion*

We emphasize that the purely abstract notion of zero-point fields, as considered by Casimir and the zero-temperature limit of Lifshitz's finite temperature theory are equivalent. Using the latter theory we have calculated numerically the temperature effect on the attractive force between perfectly reflecting mirrors. In this way we have shown that, according to Lifshitz's approach, the Casimir limit represents the highest possible value of the force.

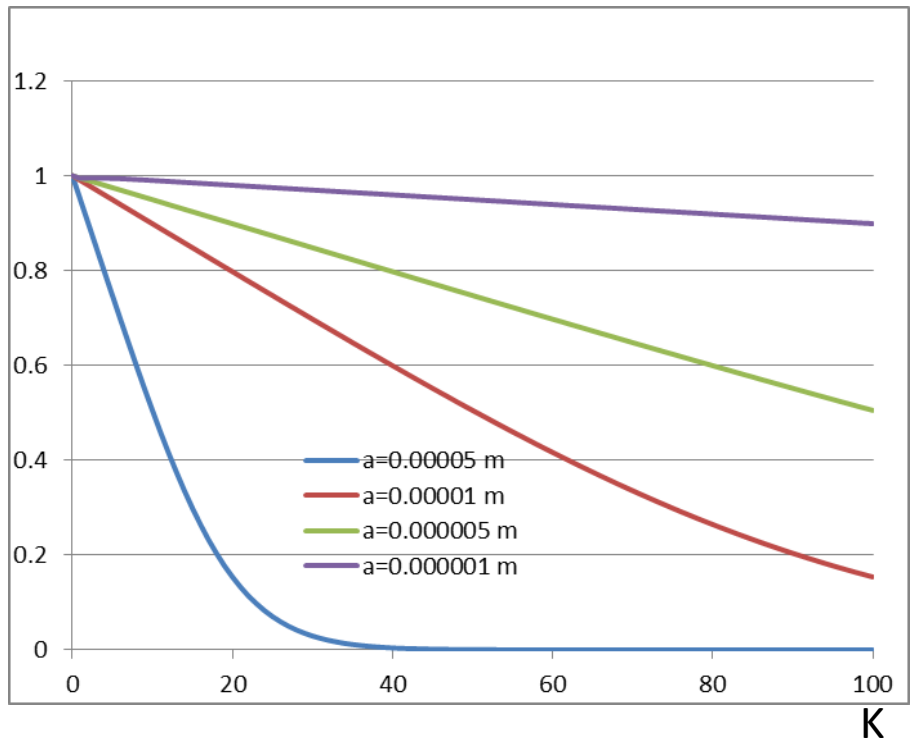

*Appendix*

The Lifshitz force

Here we summarize the calculation of Lifshitz for the special case of perfectly reflecting plates.

Using Matsubara Green's functions $D_{lk}(\varsigma_s; \mathbf{r}, \mathbf{r}')$ with $\varsigma_s$ a pseudo frequency satisfying the relation

(A1) $\hbar\varsigma_s = 2\pi s T$ with s an integer and T the temperature in energy units, he starts from the differential equation

(A2) $\left[\dfrac{\partial^2}{\partial x_i \partial x_l} - \delta_{il}\Delta + \dfrac{\varsigma_s^2}{c^2}\varepsilon(\vec{r})\delta_{il}\right] D_{lk}(\varsigma_s; \mathbf{r}, \mathbf{r}') = -4\pi\delta(\mathbf{r} - \mathbf{r}')$.

The force per unit area between dielectric plates located at distances x = 0 and x = a is then represented by the $o_{xx}$ component of the Maxwell stress tensor related to the Green's functions by the expression

(A3)

$F = \sigma_{xx}(a) =$

$\dfrac{T}{4\pi} \sum\limits_{s=0}^{\infty} \left\{ D_{yy}^E(\varsigma_s; a, a) + D_{zz}^E(\varsigma_s; a, a) - D_{xx}^E(\varsigma_s; a, a) + D_{xx}^H(\varsigma_s; a, a) + D_{zz}^H(\varsigma_s; a, a) - D_{xx}^H(\varsigma_s; a, a) \right\}$

(Here we use the notations of ref. [4] e.g. x the coordinate normal to the plates)

Note that Einstein's summation rule over repeated indices is applied throughout this text.

Once the functions $D_{lk}$ are known, one has for the electric part $D_{lk}^E$ and the magnetic part $D_{lk}^H$ the relations

(A4) $\begin{aligned} D_{lk}^E(\varsigma_s; \mathbf{r}, \mathbf{r}') &= -\varsigma_s^2 D_{lk}(\varsigma_s; \mathbf{r}, \mathbf{r}') \\ D_{lk}^H(\varsigma_s; \mathbf{r}, \mathbf{r}') &= (\nabla\times)i(\nabla'\times)_{km} D_{lm}(\varsigma_s; \mathbf{r}, \mathbf{r}') \end{aligned}$

In order to solve the general equations (A2) one uses the transformation

(A5) $D_{lk}(\varsigma_s; \mathbf{r}, \mathbf{r}') = \dfrac{1}{(2\pi)^2} \int D_{lk}(\varsigma_s; q, x, x') e^{iq(y-y')} d^2q$

which has to be adapted to proper boundary conditions.

In the case of perfect mirrors one has inside the gap $\varepsilon(\mathbf{r}) = 1$ and at the borders

$D_{lk} = \dfrac{d}{dx} D_{lk} = 0$

**Moreover, to ensure convergence, one has to substract from the solutions compatible with the boundaries those that would have been obtained without this constraint.**

Under these circumstances one finds

(A6)
$$D_{xx} = -\frac{q^2 4\pi}{w \varsigma_s^2 \Delta} \cosh w(x-x'); \quad D_{yy} = \frac{w 4\pi}{\varsigma_s^2 \Delta} \cosh w(x-x'); \quad D_{zz} = \frac{4\pi}{w \Delta} \cosh w(x-x');$$

$$D_{xy} = D_{yx} = -\frac{4\pi i q}{\varsigma_s^2 \Delta} \sinh w(x-x')$$

with

(A7a) $\quad w = \left( \varsigma_s^2 + q^2 \right)^{1/2} \quad$ and (A7b) $\quad \Delta = 1 - e^{2wa}$

For the magnetic components one finds according to the definition (A4)

(A8)
$$D_{xx}^H = q^2 D_{zz}; \quad D_{yy}^H = \frac{\partial}{\partial x}\frac{\partial}{\partial x'} D_{zz}; \quad D_{zz}^H = q^2 D_{xx} + \frac{\partial}{\partial x}\frac{\partial}{\partial x'} D_{yy} - iq \frac{\partial}{\partial x'} D_{xy} + iq \frac{\partial}{\partial x} D_{yx}$$

Note that in these intermediate steps we set $\hbar = c = 1$.

Expliciting these expressions for $x = x' = a$ and introducing them into the equation (A3) for the force, we obtain after transforming back from q space to **r** space the result

(A9) $\quad F = -\frac{T}{\pi} \sum_s \int_0^\infty \frac{w}{\Delta} q \, dq$

Reintroducing the constants $\hbar$ and c but keeping T in energy units we first have

(A10) $\quad w^2 = \frac{\varsigma_s^2}{c^2} + q^2$

With the change of variables $q^2 = \frac{\varsigma_s^2}{c^2}(p^2 - 1)$ eq. (A9) takes the form

(A11) $\quad F = \frac{2T}{\pi} \sum_s \int_1^\infty \frac{\varsigma_s^3}{c^3} \frac{p^2 \, dp}{e^{2\frac{\varsigma_s}{c}pa} - 1}$

At the zero temperature limit the series $\varsigma_s$ is infinitely dense so that we may set

$\partial s = 1 \to \frac{\hbar}{2\pi T} d\varsigma \quad$ and replace the sum by an integral thus yielding the expression

(A12) $$F = \frac{\hbar}{\pi^2} \frac{1}{c^3} \int_0^\infty \varsigma^3 d\varsigma \int_1^\infty \frac{p^2 dp}{e^{2\frac{\varsigma}{c}pa} - 1}.$$

With $X = 2\frac{\varsigma}{c}pa$ we thus arrive at the result

(A13) $$F = \frac{\hbar c}{16\pi^2} \frac{1}{a^4} \int_0^\infty \frac{X^3}{e^X - 1} dX = -\frac{\pi^2}{8} \hbar c B_4 = \frac{\hbar c}{a^4} \frac{\pi^2}{240} \quad \text{with } B_4 = -\frac{1}{30} \text{ a Bernoulli}$$

number.

In this way the Casimir result appears as the zero temperature limit of the Lifshitz force between perfect mirrors.